\documentstyle[prd,preprint,aps,epsfig,floats]{revtex}
\begin{document}
%%%%%%%%%%%%%%%%%%%%%
%title page
%%%%%%%%%%%%%%%%%%%%%

\tightenlines
\input epsf.tex
\def\DESepsf(#1 width #2){\epsfxsize=#2 \epsfbox{#1}}
\draft
%\begin{titlepage}
\thispagestyle{empty}
\preprint{\vbox{\hbox{OSU-HEP-01-09} \hbox{UMDPP-02-032}
\hbox{January 2002}}}

\title{\Large \bf Predictive Schemes for Bimaximal Neutrino Mixings}

\author{\large\bf K.S. Babu$^{(a)}$ and R.N. Mohapatra$^{(b)}$}

\address{(a) Department of Physics, Oklahoma State University\\
Stillwater, OK 74078, USA \\}

\address{(b) Department of Physics, University of Maryland\\
College Park, MD 20742, USA}

\maketitle

\thispagestyle{empty}

\begin{abstract}

We present simple and predictive scenarios for neutrino masses which lead
to a bimaximal pattern of mixings that apparently provides the best fit
to the current solar and atmospheric neutrino data.
It is shown that an approximate
global $L_e-L_\mu-L_\tau$ symmetry, broken only by Planck scale effects,
can naturally
explain the bimaximal mixing pattern.
In one scenario, the solar neutrino oscillation
parameters are induced entirely through radiative renormalization group effects.
A one parameter model describing all neutrino data is presented, which predicts
$U_{e3}\simeq \sqrt{m_e/m_\tau} \simeq 0.017$. A second two parameter scheme
predicts $U_{e3}$ to be vanishingly small.    We also show how these scenarios
can be extended to accommodate the LSND observations within a 3+1 neutrino
oscillation scheme by including a mirror sector of particles and forces.

\end{abstract}

\section{Introduction}

There appear to be persuasive arguments in favor of the possibility that
both the solar and the atmospheric neutrino anomalies are explained by a
specific pattern of neutrino mixings where both the $\nu_e-
\nu_{\mu}$ as well as the $\nu_{\mu}-\nu_{\tau}$ mixing angles are nearly
maximal,
while the $\nu_e-\nu_\tau$ mixing angle is small
\cite{solar,review,atm,chooz}. The
$\nu_e-\nu_\mu$ mixing explains the solar neutrino deficit whereas the
$\nu_\mu-\nu_\tau$
mixing accounts successfully for the atmospheric neutrino data. This
pattern has been
called in the literature as bimaximal mixing\cite{bimax}. The best fit
values for the mass difference squares for the atmospheric and solar data
seem to be $\Delta m^2_{atm}\simeq 3\times 10^{-3}$ eV$^2$ and $\Delta
m^2_{\odot}\sim (10^{-4}-10^{-5})$ eV$^2$.

A major theoretical challenge is to provide an
understanding of this pattern in reasonable extensions of the
standard model (SM). There are many models in the
literature \cite{models}, which provide ways to arrive at the bimaximal
pattern starting with different assumptions. In this note, we
suggest several simple scenarios which are quite predictive and therefore
testable in the near future in ongoing and proposed neutrino oscillation
experiments.

Before proceeding to present our schemes, we wish to motivate some special
forms of the light neutrino
mass matrix that lead to bimaximal
mixings.  These matrices are obtained under some restricted but well motivated
set of assumptions, mindful of possible approximate symmetries that might
be present in the lepton sector.  Once identified, controlled breaking of
these approximate symmetries can lead to very predictive schemes for
neutrino mixings, as we shall demonstrate.

We work in a basis where the charged lepton mass matrix is diagonal.
We define the neutrino mixing matrix (the MNS matrix) $U$ as:
\begin{eqnarray}
\left(\begin{array}{c}\nu_e\\\nu_{\mu}\\ \nu_{\tau}\end{array}\right)~=~U
\left(\begin{array}{c}\nu_1\\\nu_2\\ \nu_3\end{array}\right)
\end{eqnarray}
where the states on the left and the right hand sides denote flavor
and mass eigenstates.
The light neutrino mass matrix ${\cal M}_\nu$ is given by the equation
\begin{eqnarray}
{\cal M}_\nu~=~U{\cal M}_d U^T
\end{eqnarray}
where ${\cal M}_d$ is diagonal and has the form ${\cal M}_d~=~ Diag (m_1,
m_2, m_3)$. For the mass eigenvalues,
there are essentially three different possibilities of
physical interest: (a) hierarchical form where $m_1 \ll m_2 \ll m_3$;
(b) inverted form where $|m_1|\simeq |m_2| \gg |m_3|$ and (c) degenerate
form where $m_1=m; ~m_2=m+\delta_2;~ m_3~=~m+\delta_3$, with $\delta_{2,3}
\ll m$.
Since experiments indicate that the $\nu_\mu-\nu_\tau$
mixing angle is nearly maximal, and that $U_{e3}$ is very small, we choose
the following approximate form for the mixing matrix $U$:
\begin{eqnarray}
U~=~\left(\begin{array}{ccc} c &
s & 0\\ \frac{s}{\sqrt{2}}
& -\frac{c}{\sqrt{2}} & \frac{1}{\sqrt{2}}\\ \frac{s}{\sqrt{2}}
& -\frac{c}{\sqrt{2}} & -\frac{1}{\sqrt{2}}\end{array}\right).
\end{eqnarray}
Here $s \equiv \sin\theta$ is the solar neutrino oscillation angle, which
is of order one, while the corresponding angle for atmospheric neutrino
oscillation has been set to $1/\sqrt{2}$.  In the exact bimaximal limit,
we will have $s=c=1/\sqrt{2}$.

Consider case (a) in the exact bimaximal limit.  Let us denote $m_3 = m$,
$m_2 = \delta$ and set $m_1 = 0$.  The matrix ${\cal M}_\nu$ is then give by
\begin{eqnarray}
{\cal M}_\nu^{(a)}~=~ \left(\begin{array}{ccc} \delta/2 &
-\delta/(2\sqrt{2}) &-\delta/(2\sqrt{2}) \\
-\delta/(2\sqrt{2}) & m/2+\delta/4 &
\delta/4-m/2\\ -\delta/(2\sqrt{2})
&\delta/4-m/2 & \delta/4+m/2
\end{array}\right)~.
\end{eqnarray}
For case (c), the form is the same as Eq. (4), except that
there is an additional piece proportional
to the unit matrix, $Diag(1,1,1)m_0$, with $m_0$ setting
the overall scale for the degenerate
mass. The relative hierarchy among the parameters in this case will be
$m_0\gg m \gg \delta$.

For case (b) with the inverted spectrum, we get
(choosing for simplicity the case $c=s =\frac{1}{\sqrt{2}}$ and $m_3=0,~
m_1=m=-m_2+\delta$),
\begin{eqnarray}
{\cal M}_\nu^{(b)} =
\left(\begin{array}{ccc} \delta/2 & (2m-\delta)/(2\sqrt{2}) &
(2m-\delta)/(2\sqrt{2}) \\ (2m-\delta)/(2\sqrt{2}) &
\delta/4 & \delta/4
\\(2m-\delta)/(2\sqrt{2}) & \delta/4 & \delta/4
\end{array}\right)~.
\end{eqnarray}
Of these forms, Eq. (5) is particularly interesting since if we set
$\delta=0$ it reduces to
\begin{eqnarray}
{\cal M}_\nu^{(0)} = \left(\begin{array}{ccc} 0 & m & m\\ m & 0 & 0 \\
m & 0 & 0\end{array}\right)~
\end{eqnarray}
with a redefined $m$.
The mass matrix given in
Eq. (6) has the interesting property that it has a global symmetry
$L_e-L_{\mu}-L_{\tau}$ ($L_i$ is the $i$th lepton number) along with a
permutation symmetry $S_2$ acting on the $\nu_{\mu}, \nu_{\tau}$ flavors.
This new symmetry unraveled in the leptonic world
is perhaps analogous to the local $B-L$ symmetry that is revealed
by the seesaw mechanism \cite{seesaw} for neutrino masses.

As it stands, ${\cal M}_\nu^{(0)}$ of Eq. (6)
cannot lead to solar neutrino oscillations since $\Delta m^2_{\odot}=0$.
One needs to break the $L_e-L_\mu-L_\tau$ symmetry in order to make the
model realistic.
Since the mass parameter $m$ in Eq. (6) is fixed by the atmospheric
data i.e., $m\simeq \sqrt{\Delta m^2_{atm}}$, the strength of the
$L_e-L_{\mu}-L_{\tau}$ breaking should be of order $\Delta
m^2_{\odot}/\Delta m^2_{atm}\approx (3\times 10^{-2}-3\times 10^{-3})$.
The smallness of this symmetry breaking parameter suggests that
$L_e-L_\mu-L_\tau$ as well as $S_2$ symmetries are indeed approximately
conserved.

In this paper, we attempt to seek realistic gauge models that realize
Eq. (6) as the
leading approximation of the light neutrino mass matrix.  The correction terms
to the mass matrix will involve very few parameters, so that
there is a certain predictive power for neutrino mass difference squares
as well as
mixings. The hope is that such models can then be tested in future
experiments and may shed light into the
nature of new physics operating in the neutrino sector and
perhaps more generally within quark lepton unified schemes. We present several
examples of such models. First we show that global $L_e-L_\mu-L_\tau$ symmetry,
broken only by Planck scale effects, can give a rather good fit to the neutrino
data.  Then we present a scheme where the neutrino sector has only
one parameter which is fixed by the atmospheric neutrino data. Radiative
renormalization effects from the charged lepton sector then determines
the rest of the parameters in the neutrino mass matrix, making the model
extremely predictive. For instance, in this model, we obtain the relation
\begin{eqnarray}
\Delta m^2_{\odot}\simeq \Delta m^2_{atm}
4\sqrt{2}r\sqrt{\frac{m_e}{m_{\tau}}}
\end{eqnarray}
where $r$ is a calculable parameter which depends only on the scale of new physics
(e.g. seesaw
scale) governing the smallness of $m_{\nu}$. In this model, the mixing
parameter $U_{e3}$ is predicted to be $U_{e3}\simeq
\sqrt{\frac{m_e}{m_{\tau}}}\simeq 0.017$. We also present a variation of this
model with one more parameter, where $U_{e3}$ can be larger (even
close to the present experimental upper limit of $0.2$).
We present a second scenario which has two parameters.
This model predicts bimaximal
neutrino mixings with $U_{e3}=0$. Thus
any evidence for nonzero $U_{e3}$ would rule out this scenario.
Finally, we consider extensions of these schemes that may
provide a
simultaneous fit to solar, atmospheric as well as the LSND neutrino oscillation
data.  As is well known, this needs the introduction of a
sterile neutrino. We achieve this in a manner that introduces only
two more parameters into the theory by invoking the
mirror matter models, where all new parameters (except the weak scale and
possibly the QCD scale) are fixed by an exact mirror
symmetry \cite{lee,mirror}.

\section{$L_e-L_{\mu}-L_{\tau}$ symmetry and bimaximal mixing}
Suppose that $L_e-L_\mu-L_\tau$ symmetry is a global symmetry of the
 Lagrangian broken only by Planck scale induced terms.  (Quantum
gravity is expected to break all non gauge symmetries.)
The Lagrangian in this case will have the general form:
\begin{eqnarray}
{\cal L} ~=~ \frac{1}{M} (a L_e\phi{L_{\mu}}\phi
+b L_e\phi{L_{\tau}}\phi) +
\frac{f_{\alpha \beta}}{M_{P\ell}}(L_{\alpha}\phi L_{\beta}\phi)+ h.c.
\end{eqnarray}
where $\alpha,\beta$ are flavor indices. Here $\phi$ is the standard model
Higgs doublet.  $M$ is the seesaw scale,
which can be near the GUT scale $\sim 10^{15}$ GeV, so the first
set of terms in Eq. (8) will
dominate over the last set.  Note that the $L_e-L_\mu-L_\tau$ symmetry
is broken by the Planck scale induced terms.  The leading
contribution to ${\cal M}_\nu$ is of the form in Eq. (6), with the
two nonzero entries unequal in general (denote them $m_1$ and $m_2$).
Upon diagonalizing, one obtains two nearly degenerate neutrino
eigenstates with opposite CP parities with a common mass of
$m=\sqrt{m_1^2+m_2^2}$.  If $M$ is near the GUT scale, $m \sim 0.03$ eV
can be explained quite naturally.  A bimaximal pattern of neutrino mixings will
also be induced, with $U$ given as in Eq. (3).

At this stage, we cannot account for solar neutrino oscillation since
$\Delta m^2_{\odot}=0$ from Eq. (6). Once we include the Planck scale
corrections from the last term of Eq. (8), one
gets a mass matrix
\begin{eqnarray}
{\cal M}_{\nu}~=~\left(\matrix{\delta & m_1 & m_2 \cr m_1 & \epsilon_1 &
\epsilon_2 \cr m_2 & \epsilon_2 & \epsilon_3}\right)
\end{eqnarray}
with $\delta, \epsilon_{i}\simeq 2\times 10^{-5}$ eV.
This leads to
$\Delta m^2_{\odot}\simeq 10^{-6}$ eV$^2$, which is close to the required
value for large mixing angle solar neutrino oscillation fit. It is
encouraging that all the
parameters are in the right range to fit solar and atmospheric neutrino
oscillation given the assumption that the seesaw scale $M$ is near the GUT
scale.  It should be noted that in this scenario, the solar oscillation
angle is given by $\sin^22\theta_\odot = 1-{\cal O}(\Delta m^2_\odot/\Delta 
m^2_{atm})^2$, which is very close to 1.  Thus this scenario would prefer the
LOW solution to the solar neutrino puzzle, although the LMA solution is not
entirely excluded.  As we shall see in the next section, small breaking
of $L_e-L_\mu-L_\tau$
symmetry in the charged lepton sector would make this scenario
nicely consistent with the LMA solution.  

We now turn to making this type of models more predictive.

\section{ Radiatively induced solar neutrino oscillations}
In this section, we present a single parameter model for neutrino
masses where the solar neutrino oscillations are generated by radiative
renormalization group effects. We will provide a gauge theoretic
derivation of the model in section IV.
Consider the following neutrino mass matrix defined at a (unification)
scale much above the weak scale as in Eq. (6).
\begin{eqnarray}
{\cal M}_\nu^0 = \left(\matrix{0 & 1 & 1 \cr 1 & 0 & 0 \cr 1 & 0 & 0}\right) m~.
\end{eqnarray}

\noindent Let us assume that the charged lepton mass matrix in the same
basis is of the form:
\begin{eqnarray}
{\cal M}_{\ell^+} = \left(\matrix{ 0 & 0 & x \cr 0 & y & 0 \cr x' & 0 &
1}\right) m_\tau~.
\end{eqnarray}
For the most part we will take $x'=x$ so that $|x| \simeq \sqrt{m_e/m_\tau}$
and $y \simeq m_\mu/m_\tau$.  We will also consider the possibility that
$x \gg x'$ (the right handed singlet leptons multiply the matrix in
Eq. (11) on the right). In the case $x=x'$, note that there is only parameter
($m$ of Eq. (10)) in the
leptonic sector (both the charged leptons and neutrinos). We will show
that this model can lead to a realistic description of the neutrino
oscillations.

Below the unification scale, through the renormalization of
the effective $d=5$ neutrino mass operator the form of
$M_\nu^0$\cite{babu} will be modified.
(Analogous corrections in $M_{\ell^+}$ is negligible.)
The modified neutrino mass matrix at the weak scale is given by

\begin{eqnarray}
{\cal M}_\nu \simeq {\cal M}_\nu^0 + {c \over 16 \pi^2} {\rm ln}(M_U/M_Z)\left(
Y_\ell Y_\ell^\dagger M_\nu^0 + M_\nu^0 (Y_\ell Y_\ell^\dagger)^T\right)~.
\end{eqnarray}
Here $c=-3/2$ for SM while $c=1$ for SUSY, and $Y_{\ell}$ is the
charged lepton Yukawa coupling matrix.
We have absorbed the flavor--independent renormalization factor into
the definition of $m$ in Eq. (10).  Explicitly,

\begin{eqnarray}
{\cal M}_\nu \simeq \left(\matrix{2 r x & 1+r(x^2+y^2) & 1+r(1+2x^2) \cr
1+r(x^2+y^2) & 0 & rx \cr
1+r(1+2x^2) & rx & 2rx}\right)m ~,
\end{eqnarray}
where
\begin{equation}
r \equiv {c \over 16 \pi^2} Y_\tau^2 {\rm ln}(M_U/M_Z)~.
\end{equation}
Numerically, $r\simeq -3.7 \times 10^{-5}$ in the SM, while it
is $r \simeq 0.022 (\tan\beta/30)^2$ in SUSY.  The parameters
$x \simeq 0.017$ and $y \simeq 0.059$ are also much smaller than
1, facilitating an expansion in small $[r,x,y]$.  Note that in
the case of $x' \neq x$, the numerical value of $x$ can be larger
than $0.017$ by a factor even as large as 10.  (The $e-\tau$ mixing
will be still small if $x\simeq 0.17$, and will not upset the
success of approximate $L_e-L_\mu-L_\tau$ symmetry.)

The eigenvalues of $M_\nu$ to leading order in $(r,x,y)$ are
(in units of $m$)

\begin{eqnarray}
m_1 &\simeq& \sqrt{2} + {r \over 2}(4x + \sqrt{2})+
{r^2 \over 8}\sqrt{2} \nonumber \\
m_2 &\simeq& -\sqrt{2}-{r \over 2}(-4x+\sqrt{2})
-{r^2 \over 8}\sqrt{2} \nonumber \\
m_3 &\simeq& -r^2x~,
\end{eqnarray}
where terms of order $r^3, x^3, y^3$ etc have been dropped.

The leptoninc mixing matrix $U$ is obtained from $U = V_{\ell} V_\nu^T$, where
$V_{\ell}$ diagonalizes the charged lepton matrix and $V_\nu$ the neutrino
mass matrix:
\begin{eqnarray}
U=\left[\matrix{{1 \over \sqrt{2}} - {x \over 2}(1 + {r \over 2})
& -{1 \over \sqrt{2}} - {x \over 2}(1 + {r \over 2}) &
-{r x \over \sqrt{2}} -{x \over \sqrt{2}} (1 - {r \over 2}) \cr
{1 \over 2} - {r \over 4}(1+\sqrt{2}x)+{r^2 \over 16} &
{1 \over 2} + {r \over 4}(-1 +\sqrt{2}x) & -{1 \over \sqrt{2}} - {r \over
2\sqrt{2}} + {3 \over 8 \sqrt{2}} r^2 \cr
{1\over 2} + {x \over \sqrt{2}} + {r \over 4}(1 + \sqrt{2} x)-{3 \over
16}r^2 & {1\over 2} -{x \over \sqrt{2}} +{r\over 4}(1-\sqrt{2} x)-{3
\over 16} r^2 & {1 \over \sqrt{2}} -{r \over 2\sqrt{2}} + {r^2 \over 8\sqrt{2}}
} \right]~.
\end{eqnarray}

$U$ has approximately the bimaximal form (compare Eq. (16) with Eq. (3)).
Let us identify $m_1$ and $m_2$ to have masses of order 0.05 eV ($\sqrt{2}m \simeq 0.05
$eV) so that
the atmospheric neutrino oscillation is explained via $\nu_\mu-\nu_\tau$
oscillations.  The solar mass splitting is then given by
$\Delta m^2_{\odot} = m_1^2 -m_2^2 \simeq 8 \sqrt{2} r x m^2 \simeq
0.014 rx$ eV$^2$.  In the SM, this is equal to $0.92 \times 10^{-8}$ eV$^2$
if $x=x'$.  On the other hand, if $x = 10 x'$, $\Delta m^2_{\odot}
\simeq 0.92 \times 10^{-7}$ eV$^2$, which is in the right range for LOW
solution of the solar neutrino data.  In SUSY, even with $x=x'$, we get
$\Delta m^2_{\odot} \simeq 4.7 \times 10^{-6}$ eV$^2$ for $\tan\beta
= 30$, which is the right range for LMA solution.  In the case of
SUSY with $\tan\beta = 30$, if we choose $x=0.2$, so that $U_{e3} \simeq 0.15$,
which is near its experimental upper limit, we have
$\Delta m^2_{\odot} \simeq
5.5 \times 10^{-5}$ eV$^2$ and the solar mixing angle is given by
$\sin^22\theta_{\odot} \simeq 0.88$.
These values are nicely consistent with the large mixing
angle MSW solution.

Neutrinoless double beta decay occurs in this model with an amplitude
proportional to (see Eq. (16))
\begin{equation}
m^{\rm effective}_{\beta\beta0\nu} \simeq \left({ 1 \over \sqrt{2}}- {x \over
2}\right)^2
m_1 +\left(-{1 \over \sqrt{2}} - {x \over 2}\right)^2
m_2 \simeq \sqrt{2}x m_1 \simeq
0.05~x ~ {\rm eV}.
\end{equation}
For $x= x' \simeq 0.017$, this is $\simeq 10^{-3}$ eV, while for
$x = 10 x'$, this is 0.01 eV.

\section{A gauge model}

The model for neutrino masses described in the previous section can be
derived by extending the standard model to include extra gauge singlets
$S_0\equiv (S_1,S_2)$ which transform as a doublet under an $S_3$
permutation group \cite{pakvasa}.
The left handed lepton doublets $\tilde{L}\equiv (L_2, L_3)$ of the second
and third
generation are assumed to transform also as doublets of $S_3$. The right
handed lepton fields as well as all other standard model fields are assumed
to be singlets under $S_3$. To obtain the desired charged lepton mass matrix
we include two more pairs of standard model singlets, $S_{u,d}$ each
transforming as a doublet under the $S_3$ group. We also assume a $Z_2$
symmetry under which $S_d, \tau_R$ are odd and all other fields are even.

The most general $L_e-L_{\mu}-L_{\tau}$ and $S_3$ invariant coupling of
leptons is given by:
\begin{eqnarray}
{\cal L}_1 ~=~ \frac{1}{M^2} L_e\phi\tilde{L}\phi S_0 + \frac{1}{M}
\bar{\tilde{L}}\phi S_d \tau_R + \bar{\tilde{L}}\phi S_u \mu_R  + h.c.
\end{eqnarray}
Here $M$ is a fundamental scale close to the GUT scale.  The singlets
$S_0, S_u,S_d$ all acquire VEVs of order $M$.  The $S_3$ invariant
potential for these fields will admit the following structure of VEVs:
$\left\langle S_0 \right\rangle = v_0~(1,1),~\left\langle S_u \right
\rangle = v_u~(1,0)$ and $\left\langle S_d \right\rangle = v_d~(0,1)$
\cite{bb}.  Such a VEV structure will lead to
the charged lepton and neutrino mass matrices given
in Eq. (10)-(11) except for the fact that $x=x'=0$. To induce $x,x'\neq 0$, we
need to introduce singlet Higgs fields that break the
$L_e-L_{\mu}-L_{\tau}$ symmetry. This can also be done in simple ways.

\section{A two parameter model from permutation symmetry}
In this section, we will present a two parameter model that can also
emerge from the model of the previous section if we allow for terms that break
$L_e-L_{\mu}-L_{\tau}$ symmetry.
Consider the following Lagrangian:
\begin{eqnarray}
{\cal L}_1 ~=~ \frac{1}{M^2} L_e\phi\tilde{L}\phi S_0 +
\frac{1}{M}(L_{\mu}\phi L_{\mu}\phi+L_{\tau}\phi L_{\tau}\phi) +
\tilde{\bar{L}}\phi S_d \tau_R + \tilde{\bar{L}}\phi S_u \mu_R  + h.c.
\end{eqnarray}
In this case, one obtains the Majorana neutrino mass matrix to be
\begin{eqnarray}
{\cal M}_\nu^0 = \left(\matrix{0 & 1 & 1 \cr 1 & a & 0 \cr 1 & 0 & a}\right) m~.
\end{eqnarray}
The charged lepton mass matrix is diagonal here.
This mass matrix can be diagonalized to give the neutrino mixing matrix
$U$, exactly as in Eq. (3).
The solar neutrino
oscillation is governed by $\Delta m^2_{\odot} \simeq 2m^2 a$.  This model
predicts $U_{e3}=0$ as can be seen from Eq. (3).
Thus any evidence for nonzero $U_{e3}$ would rule out this model.

\section{Mirror neutrinos and an explanation of LSND experiment}
So far we have ignored the LSND results\cite {lsnd} in our discussion of
the positive neutrino oscillation signals. In order to accommodate the
LSND results, one will have to include an extra neutrino species that is a
standard model singlet (a sterile neutrino).
In the presence of a sterile neutrino, there are two ways to
understand the observations: one is the so--called 2+2
scheme \cite{cald} and the other is the 3+1 scheme \cite{3p1}.
The 2+2 scheme has the $\nu_{\mu,\tau}$ neutrinos with mass
around an eV and $\nu_{e,s}$ with mass near $10^{-3}$ eV, with the latter
explaining the solar neutrino data, the former explaining the
atmospheric neutrino data and the gap between the two pairs explaining
the LSND results. Recent SNO data disfavors the original version
of the 2+2 model where all the missing solar $\nu_e$'s are converted via
a small angle MSW mechanism only to the sterile neutrinos. Mixed 2+2
scenarios where there is substantial mixing between the two sectors
separated by a gap seem to be consistent with all data \cite{mixed}.

In the 3+1 picture, on the other hand, it is assumed that the three
active neutrinos are bunched together at a small mass value (say around
$6\times 10^{-2}$ eV or so), with the sterile neutrino at a mass near
an eV. The atmospheric and solar neutrino data is explained by the
oscillations among active neutrinos whereas the LSND data is explained by
indirect oscillations involving the sterile neutrino\cite{pati}. The
advantage of
this picture in view of the SNO data is that the flat energy spectrum is
explained by postulating bimaximal neutrino mixing pattern among
the active neutrinos or the LOW solution, as discussed in previous
sections of this paper.
The question now is whether one can extend the model we analyzed
to include one or more light sterile neutrinos and maintain at
least some predictive
power that we had in the three neutrino case.

One way to obtain light sterile neutrinos is to invoke a mirror
sector for the particles and forces, as an exact replica of the
standard model sector.  One may then identify
the neutrinos of the mirror sector as sterile
neutrinos. The ultralightness of the sterile neutrinos is then
explained in a simple way by a mirror version of the seesaw mechanism. In this
picture, the two sectors are connected only by gravity or possibly
by very heavy particles that decouple from low energy physics. The mixing
between neutrinos then owes its origin either to gravity or to operators
generated by the exchange of the heavy particles that connect the two
sectors.

In such theories there are two possibilities for the weak scale for
the mirror sector: (i)
it is the same as in the familiar sector or (ii) it is different.
In order to explain the LSND results in the 3+1 scenario, we need to work
with the second scenario, where mirror weak scale $v'\gg v$.  We will
consider a mirror embedding of the model discussed in section 2 which was
based only on $L_e-L_{\mu}-L_{\tau}$ symmetry. We will assume that there
is an identical symmetry i.e., $L'_{e}-L'_{\mu}-L'_{\tau}$ in the mirror
sector and as in sec. II we will further assume that only Planck scale
effects break these two symmetries.  They are also
responsible for mixing between the two sectors. The Lagrangian
including the Planck scale breaking terms can then be written as:
\begin{eqnarray}
{\cal L} ~&=&~ \frac{1}{M} (a L_e\phi{L_{\mu}}\phi
+b L_e\phi{L_{\tau}}\phi) +
\frac{f_{\alpha \beta}}{M_{P\ell}}(L_{\alpha}\phi L_{\beta}\phi)
+ \frac{1}{M'} (a L'_e\phi'{L'_{\mu}}\phi'
+b L'_e\phi'{L'_{\mu}}\phi') \nonumber \\
&+&
\frac{f'_{\alpha \beta}}{M_{P\ell}}(L'_{\alpha}\phi' L'_{\beta}\phi')+
\frac{g_{\alpha \beta}}{M_{P\ell}}(L_{\alpha}\phi L'_{\beta}\phi')+ h.c.
\end{eqnarray}
As mentioned before, we work in a picture where the gauge symmetry
breaking is not mirror symmetric. This then allows for the
possibility that both the weak scale and the seesaw scale in the mirror
sector are very different from those in the familiar sector. Suppose
we assume that $\left\langle \phi'\right\rangle
\equiv v'\simeq 5\times 10^4$ GeV and the mirror
seesaw scale $M'\simeq 10^{9}$ GeV or so. From Eq. (21) one sees that
the two heavy neutrino states $\nu'_{1,2}$ can have masses easily
of the order of 1-10 GeV and the lightest mirror neutrino has a mass of
$\sim $ eV and can therefore be used to understand the LSND results.
The off diagonal $\nu-\nu'$ mixing terms are of order $vv'/M_{P\ell}\simeq
0.01$ eV, which is also of the right order for the purpose.

Upon integrating out the heavy neutrinos, one gets a 4$\times$4 matrix with the
following generic structure (neglecting terms smaller than $10^{-2}$ eV):
\begin{eqnarray}
{\cal M}^{(4)}~=~\left(\matrix{0 & m_1 & m_2 & \epsilon_1 \cr m_1 & 0 & 0
& \epsilon_2 \cr m_2 & 0 & 0 & \epsilon_3 \cr \epsilon_1 & \epsilon_2 &
\epsilon_3 & \delta }\right)~.
\end{eqnarray}
For $\delta \gg \epsilon_i, m_i$, this matrix has one large eigenvalue
$\approx \delta$, which mixes with $\nu_{e,\mu}$ with a mixing angle
of order $\epsilon_i/\delta$; for $\epsilon_i\approx 0.02$ eV and $\delta
\approx 0.2$ eV, the LSND results can be explained using indirect
oscillation \cite{balaji}, with the probability given by $P(\nu_e\rightarrow \nu_\mu)\sim
4|\epsilon_1 \epsilon_2/\delta^2|^2$.

\section{Cosmological implications}

The mirror sector scenario has several interesting cosmological implication.
First of all, the two heavy ($\approx $ GeV) mirror neutrinos annihilate
into the lightest mirror neutrino early on and disappear. As a result
they do not contribute to the expansion of the universe
during big bang nucleosynthesis (BBN).  The lightest mirror neutrino
will however be in equilibrium during BBN.
Thus, this
model would predict one extra neutrino species at the BBN. In principle,
this could be tested once the He$^4$ and D$^2$ abundances are known more
precisely.

A second interesting implication is that due to the heavy mirror scale,
the mirror QCD scale is $\Lambda'\approx 7$ GeV (for the
nonsupersymmetric case).  This is calculated as
follows. Mirror symmetry guarantees that the two QCD couplings are the same
at very high energies. As they evolve to lower scales, due to the heavier
masses of the mirror quarks
$\alpha'_s$ will develop a higher slope around $10^5$ GeV
corresponding to $t'$ decoupling and
become larger than $\alpha_s$; this effect gets further reinforced at the
$(b', c')$ masses. A simple calculation using $\alpha_s(M_Z)\simeq$
0.118 shows that $\alpha'_s \approx 1$ around 7 GeV. We define this as the
mirror QCD scale. Thus from the neutrino oscillations we are able
to predict the
mirror QCD scale. Using the analogy with familiar QCD then one can predict
that mirror proton mass is around 30-40 GeV.

The mirror electron mass
is $m_{e'}\sim \frac{v'}{v} m_e\approx 500$ MeV. This would make the
mirror hydrogen Bohr radius 1000 times smaller than the familiar
hydrogen atom, leading to a
completely new profile for the mirror sector of the universe.

The mirror baryons, being stable, can now become the dark matter of the
universe.  This picture is
very similar to the one advocated in Ref.\cite{mdm}, with an important
difference  that here the mirror electron is about 10-30 times heavier.
To see how this fits in with the dark matter picture \cite{other}, we note
that in this scenario, to reconcile the presence of the mirror photon with the
success of BBN, one needs to have the temperature of the mirror universe
somewhat (say a factor of 2) lower than the temperature of the familiar
sector \cite{asym}. One can then write
\begin{eqnarray}
\frac{\Omega_{B'}}{\Omega_B}\simeq
\left(\frac{T'}{T}\right)^3\frac{m_{p'}}{m_p}~.
\end{eqnarray}
For $m_{p'}/m_p\simeq 40$ and $T'/T\simeq 0.5$, we get
$\Omega_{B'}/\Omega_B \simeq 5$. If we take $\Omega_B\simeq 0.05$, this
leads to $\Omega_{DM}\equiv \Omega_{B'}\sim 25\%$, consistent with the
current thinking in the field of dark matter physics.

\section{Conclusions}

We have presented several simple and predictive three neutrino
oscillation scenarios that can explain the solar and the atmospheric
neutrino data
with a bimaximal mixing pattern that seems to be favored by current
data, especially from the flat energy
distribution for solar neutrinos. The measurement of the mixing parameter
$U_{e3}$ and evidence for inverted pattern of masses would constitute
tests of these models. We have also presented a mirror extension of the
schemes that can accommodate the LSND results as well. We find that in the simplest
mirror extension, the mirror QCD scale is predicted to be around 5-7
GeV. This leads to a mirror baryon mass mass around 40 GeV, which can then
become a viable candidate for dark matter of the universe.

\section*{Acknowledgments}

The work of KSB is supported by the DOE grant No. \#
DE-FG03-98ER-41076, a grant from the Research Corporation, and by
DOE Grant \# DE-FG02-01ER45684.
The work of RNM is supported by the National Science Foundation Grant
PHY-0099544.


\begin{thebibliography}{99}



\bibitem{solar} B.T. Cleveland et. al., {\it Astrophys. J.} {\bf 496}, 505
(1998); R. Davis, {\it Prog. Part. Nucl. Phys.} {\bf 32}, 13 (1994);
SuperKamiokande collaboration, Y. Fukuda et. al.,
{\it Phys. Rev. Lett.} {\bf 81}, 1158 (1998); Erratum $ibid$., {\bf 81},
4279 (1998) and $ibid$., {\bf 82}, 1810 (1999); Y. Suzuki, {\it Nucl.
Phys. Proc. Suppl.} {\bf B 77},  35 (1999); Y. Fukuda et. al.,
hep-ex/0103032; SAGE collaboration, J.N. Abdurashitov et. al.,
{\it Phys. Rev.} {\bf C 60}, 055801 (1991); GALLEX collaboration, W. Hampel et. al.,
{\it Phys. Lett.} {\bf B447}, 127 (1999); SNO collaboration,
Q.R. Ahmed et. al., {\it Phys. Rev. Lett.} {\bf 87}, 071301 (2001).

\bibitem{review} For theoretical analysis see:
J. N. Bahcall, M. C. Gonzales-Garcia and C. Pena-Garay,
hep-ph/0106258; A. Bandopadhyay, S. Choubey, S. Goswami and K. Kar,
hep-ph/0106264; P. Krastev and A. Y. Smirnov, hep-ph/0108177; G. L. Fogli,
E. Lisi, D. Montanino and A. Palazzo, hep-ph/0106247; M. V. Garzelli and
C. Giunti, hep-ph/0108191; M. C. Gonzales-Garcia, M. Maltoni and
C. Pena-Garay, hep-ph/0108073; V. Barger, D. Marfatia and K. Whisnant,
hep-ph/0106207.

\bibitem{atm} SuperKamiokande collaboration, Y. Fukuda et. al.,
{\it Phys. Rev. Lett.} {\bf 82}, 2644 (1999).

\bibitem{chooz} CHOOZ collaboration, M. Apollonio et. al., {\it Phys. Lett.}
{\bf B466}, 415 (1999).


\bibitem{bimax}
V. Barger, S. Pakvasa, T. Weiler and K. Whisnant, {\it Phys. Lett.}
{\bf B437}, 107 (1998);
H. Fritzsch and Z.Z. Xing, Phys. Lett. {\bf B 440} 313 (1998);
A. Baltz, A. S. Goldhaber and M. Goldhaber, {\it Phys. Rev. Lett.}
{\bf 81}, 5730 (1998);
M. Jezabek and Y. Sumino, {\it Phys.Lett.} {\bf B440}, 327 (1998);
G. Altarelli and F. Feruglio, {\it Phys. Lett.} {\bf B439}, 112 (1998);
F. Vissani, hep-ph/9708783; I. Stancu and D. V. Ahluwalia, Phys. Lett. 
{\bf B 460}, 431 (1999). 


\bibitem{models} An incomplete list:
R. Barbieri, L. Hall, A. Strumia and N. Weiner, JHEP 9812,
017  (1998);
G. Altarelli and F. Feruglio, {\it JHEP} {\bf 9811}, 021 (1998);
S.~Davidson and S.~F.~King, Phys. Lett. {\bf B445}, 191 (1998);
R. N. Mohapatra and S. Nussinov, Phys. Rev. {\bf D60}, 013002 (1999);
C.H. Albright and S.M. Barr, {\it Phys. Lett.} {\bf B461}, 218 (1999);
Y. Nomura and T. Yanagida, {\it Phys. Rev.} {\bf D59},
017303 (1999);
R. Barbieri, G. Kane, L.J. Hall, and G.G. Ross,
hep-ph/9901228;
A. Joshipura and S. Rindani, Eur.Phys.J. {\bf C14}, 85
(2000);
R. N. Mohapatra, A. Perez-Lorenzana, C. A. de S. Pires,
Phys. Lett. {\bf B474}, 355 (2000);
Q. Shafi and Z. Tavartkiladze, Phys. Lett. {\bf B482}, 145 (2000);
A. Joshipura and S. Rindani, Phys. Lett. {\bf B 494}, 114 (2000);
T. Kitabayashi and M. Yasue,
Phys. Rev. {\bf D 63}, 095002 (2001);
R. N. Mohapatra, Phys. Rev. {\bf D64}, 091301 (2001);
A. Aranda, C. Carone and R. Lebed, Phys. Rev. {\bf D65}, 013011 (2002);
S. Barr and I. Dorsner, Nucl. Phys. {\bf B585}, 79 (2000);
H. Fritzsch and Z.Z. Xing, Prog. Part. Nucl. Phys. {\bf 45}, 1 (2000);
R. Kitano and Y. Mimura, {\it Phys. Rev.} {\bf D63}, 016008
(2001); S.~F.~King and N.~N.~Singh,
Nucl. Phys.{\bf B596}, 81 (2001);
A. Gogoladze and A. Perez-Lorenzana, hep-ph/0112034;
P. Frampton, S. Glashow and D. Marfatia, hep-ph/0201008;
K.S. Babu and S.M. Barr,  Phys. Lett. {\bf B525}, 289 (2002);
S.T. Petcov, Phys. Lett. {\bf B110}, 245 (1982).

\bibitem{seesaw}  M. Gell-Mann, P. Ramond and R. Slansky, in {\it
Supergravity}, eds. P. van Niewenhuizen and D.Z. Freedman (North
Holland 1979); T. Yanagida, in Proceedings of {\it Workshop on
 Unified Theory and Baryon number in the Universe}, eds.
O. Sawada and A. Sugamoto (KEK 1979);  R.N. Mohapatra and
G. Senjanovi{\'c}, Phys. Rev. Lett. {\bf 44}, 912 (1980).


\bibitem{lee} T. D. Lee and C. N. Yang, Phys. Rev. {\bf 104}, 254 (1956);
K. Nishijima, private communication; Y. Kobzarev, L. Okun and I. Ya
Pomeranchuk, Yad. Fiz. {\bf 3}, 1154 (1966);  M. Pavsic, Int. J. T. P.
{\bf 9}, 229 (1974); S. I. Blinnikov and M. Y. Khlopov, Astro. Zh. {\bf
60}, 632 (1983); E. W. Kolb, D. Seckel and M. Turner, Nature, {\bf 514},
415 (1985); R. Foot, H. Lew and R. Volkas, Phys. Lett. {\bf B 272}, 67
(1991).

\bibitem{mirror} R. Foot and R. Volkas, Phys. Rev. {\bf D 52}, 6595
(1995); Z. Berezhiani and R. N. Mohapatra, Phys. Rev. {\bf D 52}, 6607
(1995).

\bibitem{babu}  K.S.~Babu, C.N.~Leung and J.~Pantaleone,
Phys.~Lett.~{\bf B319}, 191 (1993); P.H.~Chankowski and Z.~Pluciennik,
Phys.~Lett.~{\bf B316}, 312 (1993);
S. Antusch, M. Drees, J. Kersten, M. Lindner and M. Ratz
Phys. Lett. {\bf B519}, 238 (2001).

\bibitem{pakvasa} See e.g. H. Sugawara and S. Pakvasa, {\it Phys. Lett.}
{\bf B73}, 61 (1978).

\bibitem{bb} See eg.,  K.S. Babu and S.M. Barr, Phys. Lett. {\bf B525},
289 (2002);
For this type of VEV pattern obtained from unitary groups, see
R. Barbieri, L. Hall, G. Kane and G. Ross, hep-ph/9901228.


\bibitem{lsnd} C. Athanassopoulos {\it et al.,} Phys. Rev. {\bf C54}
(1996) 2685;
C. Athanassopoulos {\it et al.,}, Phys. Rev. {\bf C58}, 2489  (1998).


\bibitem{cald} D. O. Caldwell and R. N. Mohapatra, Phys. Rev. {\bf D 48},
3259 (1993);
J. Peltoniemi and J. W. F. Valle, Nucl. Phys. {\bf B406}, 409 (1993);
J. Peltoniemi, D. Tomassini and J. W. F. Valle, Phys. Lett. {\bf B 298},
383 (1993); for detailed analyses and other references, see
S. M. Bilenky, C. Giunti and W. Grimus, Prog. Part. Nucl. Phys. {\bf 43},
1 (1999); N. Okada and O. Yasuda, Int. J. Mod. Phys. {\bf A 12}, 3669
(1997).

\bibitem{3p1} S. M. Bilenky, C. Giunti, W. Grimus and
T. Schwetz,  Phys. Rev. {\bf D60}, 073007 (1999); B. Balatenkin,
G. Fuller,
J. Fetter and G. McGlaughlin, Phys. Rev. {\bf C 59}, 2873 (1999);
V. Barger, B. Kayser, J. Learned, T. Weiler, K. Whisnant, Phys. Lett.
{\bf B489}, 345 (2000);  O. Peres and A. Y. Smirnov, hep-ph/0011054;

\bibitem{mixed} M. C. Gonzales-Garcia, M. Maltoni and C. Pena-Garay,
hep-ph/0108073; G.L. Fogli, E. Lisi and A. Marrone, Phys. Rev. {\bf D 64},
093005 (2001); K.S. Babu and R.N. Mohapatra,
Phys. Lett. {\bf B522}, 287 (2001);
A. Bandyopadhyay, S. Choubey, S. Goswami, K. Kar hep-ph/0110307;
S. Goswami and A. Joshipura, hep-ph/1010272;
P. Roy and S. Vempati, hep-ph/0111290.


\bibitem{pati}See eg., K. S. Babu, J. C. Pati and F. Wilczek, Phys. Lett. {\bf B
359}, 351 (1995).

\bibitem{balaji} K. Balaji, A. Perez-Lorenzana and A. Yu. Smirnov, 
Phys. Lett. {\bf B509}, 111 (2001).

\bibitem{mdm} R. N. Mohapatra and V. L. Teplitz, Phys. Lett. {\bf
B462}, 302 (1999); Phys. Rev. {\bf D62}, 063506 (2000); Z. Silagadze,
Phys. At. Nucl. {\bf 60}, 272 (1997); S. Blinnikov, Astro-ph/9801015.


\bibitem{other} For a review of alternative scenarios for mirror dark matter
and a more complete list of references, see
Z. Silagadze, hep-ph/0002255.

\bibitem{asym}  Z. Berezhiani, A. Dolgov and R. N. Mohapatra, Phys.
Lett. {\bf B 375}, 26 (1996); H. Hodges, Phys. Rev.
{\bf D 47}, 456 (1993).


\end{thebibliography}
\end{document}